\documentclass[aps,pre,twocolumn,superscriptaddress,amsmath]{revtex4}
\usepackage{graphicx,amssymb,color}

\begin{document}

%\preprint{}

\title{Distributions of Bubble Lifetimes and Bubble Lengths in DNA}

\author{M.~Hillebrand} \email{malcolm.hillebrand@gmail.com} \affiliation{Department of Mathematics and
Applied Mathematics, University of Cape Town, Rondebosch 7701,
South Africa}
\author{G.~Kalosakas}  \email{georgek@upatras.gr} \affiliation{Department of Materials Science, University of Patras, GR-26504 Rio, Greece}
\author{Ch.~Skokos} \email{haris.skokos@uct.ac.za}
%\homepage{http://math\_research.uct.ac.za/~hskokos/}
\affiliation{Department of Mathematics and
Applied Mathematics, University of Cape Town, Rondebosch 7701,
South Africa}
\author{A.R.~Bishop} \email{arb@lanl.gov} \affiliation{Los Alamos National Laboratory,
Los Alamos, NM, 87545, United States}

\date{\today}

%=========================
\begin{abstract}
We investigate the distribution of bubble lifetimes and bubble lengths in DNA at physiological temperature,
by performing extensive molecular dynamics simulations with the Peyrard-Bishop-Dauxois (PBD) model,
as well as an extended version (ePBD) having a sequence-dependent stacking interaction,
emphasizing the effect of the sequences' guanine-cytosine (GC)/adenine-thymine (AT) content on these distributions.
For both models we find that base pair-dependent (GC vs AT) thresholds for considering complementary nucleotides
to be separated  are able to reproduce  the observed dependence of the melting temperature on the GC content
of the DNA sequence. Using these thresholds for base pair openings, we obtain bubble lifetime distributions
for bubbles of lengths up to ten base pairs as the GC content of the sequences is varied,
which are accurately fitted with stretched exponential functions. We find that for both models
the average bubble lifetime decreases  with increasing either the bubble length or the GC content.
In addition, the obtained bubble length distributions are also fitted by appropriate stretched exponential functions
and our results show that short bubbles have similar likelihoods for any GC content,
but longer ones are substantially more likely to occur in AT-rich sequences.
We also show that the ePBD model permits more, longer-lived, bubbles than the PBD system.
\end{abstract}
%=========================

% \pacs{05.45.-a, ????}
% \keywords{}

\maketitle

%=========================
\section{Introduction} % (fold)
\label{sec:introduction}

Over the past two decades, the study of thermally induced transient local openings in double stranded DNA
(the so-called bubbles) has given valuable insight into the potential effect of DNA dynamics on gene transcriptional
activity. The fundamentally dynamic process of transcription, which requires the opening of the DNA helix to allow
formation of the corresponding RNA strand and then closing again, has prompted the idea that DNA dynamics
may be an intrinsic factor in the very first stages of transcription \cite{sobel,Choi2004}. Bubble opening profiles of various promoter sequences have been studied extensively, revealing correlations between the transcription start site (TSS)
or other transcription factor binding sites and regions of high propensity for bubble formation
\cite{Choi2004, Kalosakas2004,Choi2008,Alexandrov2009,Alexandrov2010,Apostolaki2011,faloPRE12,huangJBE,Nowak13,faloPLOS},
suggesting that large fluctuational openings  of double stranded DNA may play some role in the process of transcription. Moreover, investigating the lifetimes of bubbles through Langevin molecular dynamics, it has been found that in several
experimentally well-studied promoters, long-lived bubbles tend to form particularly frequently at the TSS
\cite{Alexandrov2006,Alexandrov2009, Alexandrov2010}.

The advent of coarse-grained mesoscale models has been a major factor enabling the study of bubbles in DNA.
In particular, the Peyrard-Bishop-Dauxois (PBD) model \cite{PBD} has proved to be very successful in reproducing
various experimental observations. The model has been developed over time to include a nonlinear coupling
to accurately model stacking interactions between the base pairs,  resulting in the observed sharp denaturation curve
of DNA molecules \cite{Peyrard1993,DPB93,Dauxois1995,Cule1997}. This nonlinearity has also been shown to be crucial
for the formation of bubbles in double stranded DNA \cite{Theod2008}. The PBD model has been used extensively to
investigate various properties of DNA, from quantifying its chaoticity \cite{Barre2001,Hillebrand2019},
to studying signatures of localized large thermal openings in the dynamic structure factor \cite{Voulgarakis2004},
examining non-exponential decay of base pair opening fluctuations \cite{Kalosakas2006}, and more
\cite{Peyrard2000,Peyrard2004,Ares2005,nvoul,Ares2007,Kalosakas2009,weber09,falo10,Maniadis2011,Traverso2015,weber19,Muniz2020}.
Beyond this, other models have been devised to study different aspects of DNA activity
\cite{Manghi2016,barbi,cocco,metzler04,weber06,kaxiras12,zoli13,depablo14,zoli18,zoli18b,sun19}.

The importance of bubbles extends beyond studying DNA's transcriptional function, as for example the presence of bubbles has been found to impact charge transport in DNA molecules \cite{Kalosakas2003,dirk,maniadis,diaz,velarde}. Particularly the propagation of a charge along the double helix
interacts with bubble openings \cite{Kalosakas2005,Kalosakas2011,Gu2016}, while mobile discrete breathers \cite{gts} have been suggested as playing a role in charge trapping in DNA \cite{Chetverikov2019}.

In this work, considering the PBD model, as well as an extended version of it (ePBD) which takes into account the particular type of neighboring base pairs in the stacking interaction parameters, we present statistical properties of DNA bubbles,
including a detailed numerical study of the distributions of bubble lifetimes and lengths in arbitrary DNA sequences at physiological temperature ($T$=310 K).
The paper is organized as follows. In Sect.~\ref{sec:dna_models} we describe the PBD and the ePBD models used in this investigation and calculate the energy-temperature curves of the two systems.
In Sect.~\ref{sec:bubble_opening_thresholds} we suggest physical thresholds for considering base pairs to be open in the studied models and show that they are consistent with conventional melting examinations.
Then, using these thresholds, in Sects.~\ref{sec:bubble_lifetime_distributions} and \ref{sec:bubble_length_distributions} respectively we present the distributions of bubble lifetimes and bubble lengths, and discuss their characteristics. Finally, in Sect.~\ref{sec:conclusions} we summarize our results and mention some future directions for research.

%=========================
\section{DNA Models} % (fold)
\label{sec:dna_models}

In this work we use the PBD model of DNA, as well as its extended version ePBD (see below), to study DNA sequences using
microcanonical molecular dynamics. In the PBD framework, the on-site intra-base pair interactions are modeled by a Morse potential $V$,
\begin{equation}
    \label{eq:Morse}
    V(y_n) = D_n\left(e^{-a_n y_n} - 1\right)^2,
\end{equation}
with $y_n$ being the relative displacement from equilibrium of the bases within the $n$th base pair of a DNA sequence.
The nonlinear stacking interaction is accounted for by an anharmonic coupling $W$,
\begin{equation}
    \label{eq:PBDStack}
    W(y_{n},y_{n-1}) = \frac{K_{n,n-1}}{2}\left(1+\rho e ^{b(y_n + y_{n-1})}\right)\left(y_n-y_{n-1}\right)^2.
\end{equation}
Thus, considering periodic boundary conditions, the resultant Hamiltonian of a DNA sequence having in total $N$ base pairs reads
\begin{equation}
    \label{eq:PBDHamiltonian}
    H = \sum_{n=1}^N  \left[ \frac{p_n^2}{2m}+V(y_n) + W(y_n,y_{n-1}) \right],
\end{equation}
where  $p_n$ are the conjugate momenta to the canonical displacements $y_n$. The parameter values used here are taken
from fittings to melting curves of short oligonucleotides \cite{Campa1998}, which have been used extensively in previous studies
(e.g.~\cite{Choi2004,Kalosakas2004,Choi2008,Alexandrov2009,Voulgarakis2004,Ares2005,Kalosakas2006,nvoul,Ares2007,Kalosakas2009,Hillebrand2019}).
These values are $m=300$ amu for the base pair reduced mass, $D_{GC} = 0.075$ eV, $a_{GC} = 6.9$ \AA $^{-1}$ and
$D_{AT} = 0.05$ eV, $a_{AT} = 4.2$ \AA$^{-1}$ for guanine-cytosine (GC) and adenine-thymine (AT) base pairs respectively
in the Morse potential, and $K_{n,n-1} = k = 0.025$ eV/\AA$^{-2}$, $\rho = 2$, and $b = 0.35$ \AA$^{-1}$ for the stacking interaction.

In the extended ePBD model, more sensitive sequence dependence is encoded by varying the coupling constant $K_{n,n-1}$
in Eq.~\eqref{eq:PBDStack} depending on the particular succession of neighboring base pairs \cite{Alexandrov2009ePBD}.
The used, sequence dependent, coupling constants are given in Table \ref{tab:ePBDStackingConstants}, for each possible
configuration of successive base pairs. This extended model has the advantage of more accurately modelling the
experimentally observed strong effects on melting temperatures of particular base sequences \cite{Alexandrov2009ePBD},
and it has been used efficiently for {\it in silico} genetic engineering of gene promoters \cite{Alexandrov2010}.
%%%%%%%%%%%%%%%%%%%%%%
\begin{table}[tb]
    \centering
    \begin{tabular}{|c|l|l|l|l|}
    \hline
    $K_{n,n-1}$ & \multicolumn{1}{c|}{C-3$^\prime$}    & \multicolumn{1}{c|}{G-3$^\prime$}   & \multicolumn{1}{c|}{A-3$^\prime$}   & \multicolumn{1}{c|}{T-3$^\prime$}    \\ \hline
    5$^\prime$-C & 0.0192 & 0.028  & 0.025  & 0.0229 \\ \hline
    5$^\prime$-G & 0.0249 & 0.0192 & 0.019  & 0.0226 \\ \hline
    5$^\prime$-A & 0.0226 & 0.0229 & 0.0228 & 0.023  \\ \hline
    5$^\prime$-T & 0.019  & 0.025  & 0.0193 & 0.0228 \\ \hline
    \end{tabular}
    \caption{\label{tab:ePBDStackingConstants}
    Values of the ePBD stacking constants $K_{n,n-1}$ of Eq.~\eqref{eq:PBDStack}, in units of eV/\AA$^{2}$.
The rows denote the base at site $n-1$ and the columns denote the base at site $n$ in the conventional
$5^\prime$-$3^\prime$ direction. The values have been obtained from Fig.~2 of Ref.~\cite{Alexandrov2009ePBD}.}
\end{table}
%%%%%%%%%%%%%%%%%%%%%%

Our microcanonical numerical simulations were performed by using symplectic integrators, which are integration techniques
designed specifically for the efficient long-time integration of Hamiltonian systems (see e.g.~\cite{Hairer2002}).
In particular, we used the fourth order symplectic Runge-Kutta-Nystr\"om method (SRKNb6) \cite{Blanes2002},
which managed to numerically preserve the constancy of Hamiltonian
Eq.~\eqref{eq:PBDHamiltonian} (usually referred to as the system's energy) with very good accuracy,
as the relative energy error $| H(t)-H(0)|/ H(0)$ was always smaller than $10^{-6}$.
The initial conditions of our simulations were set as follows: For all $n=1, 2, \ldots, N$ the initial
base pair stretchings are $y_n=0$, while $p_n$ are randomly chosen from a normal distribution with zero mean and unit variance. Then, the $p_n$ values were uniformly scaled in order to achieve the required energy $H$, Eq~ \eqref{eq:PBDHamiltonian}, or energy density $E_N=H/N$ value. We note that in all simulations we impose periodic boundary conditions, i.e.~$y_0=y_N$, $y_{N+1}=y_1$, $p_0=p_N$ and $p_{N+1}=p_1$.

As a first step in examining the properties of the PBD and ePBD models, we investigate the relationship between
the energy density $E_N$ and the temperature $T$ for the two models. Since simulations for both systems are performed
in the microcanonical ensemble at constant energy $H$, Eq.~\eqref{eq:PBDHamiltonian}, the effective temperature
of the system is estimated using the mean kinetic energy per base pair $\langle K\rangle=\frac{1}{N}\sum_n p_n^2/(2m)$,
through the  relation $T=2\langle K \rangle/k_B$, with $k_B = 8.617 \cdot 10^{-5}$ eV/K being the Boltzmann constant.
Computing this effective temperature at different energy densities for the two models yields similar but
quantitatively slightly different behaviors.

\begin{figure}
    \centering
    \includegraphics[width=0.48\textwidth]{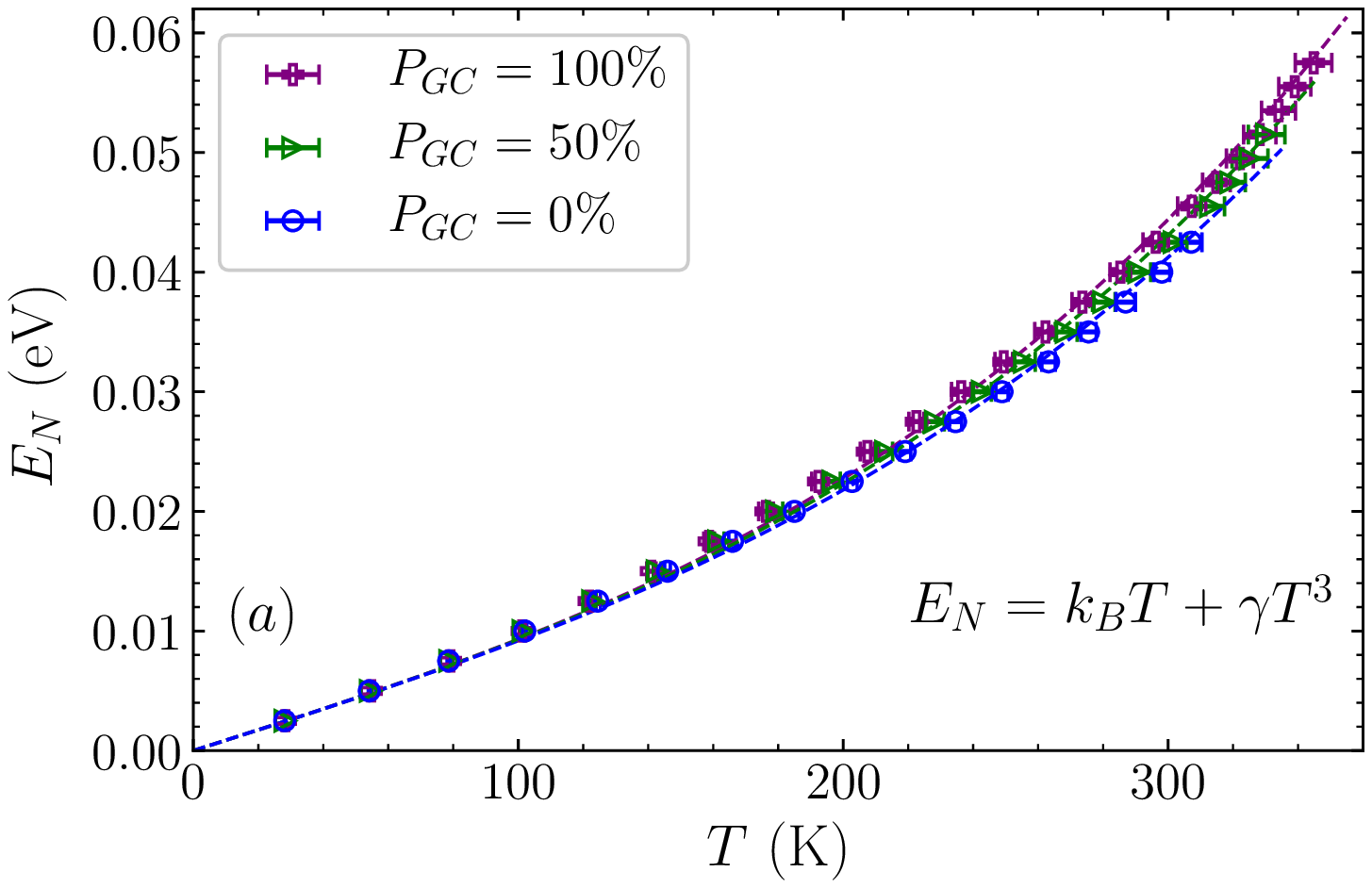}
    \includegraphics[width=0.47\textwidth]{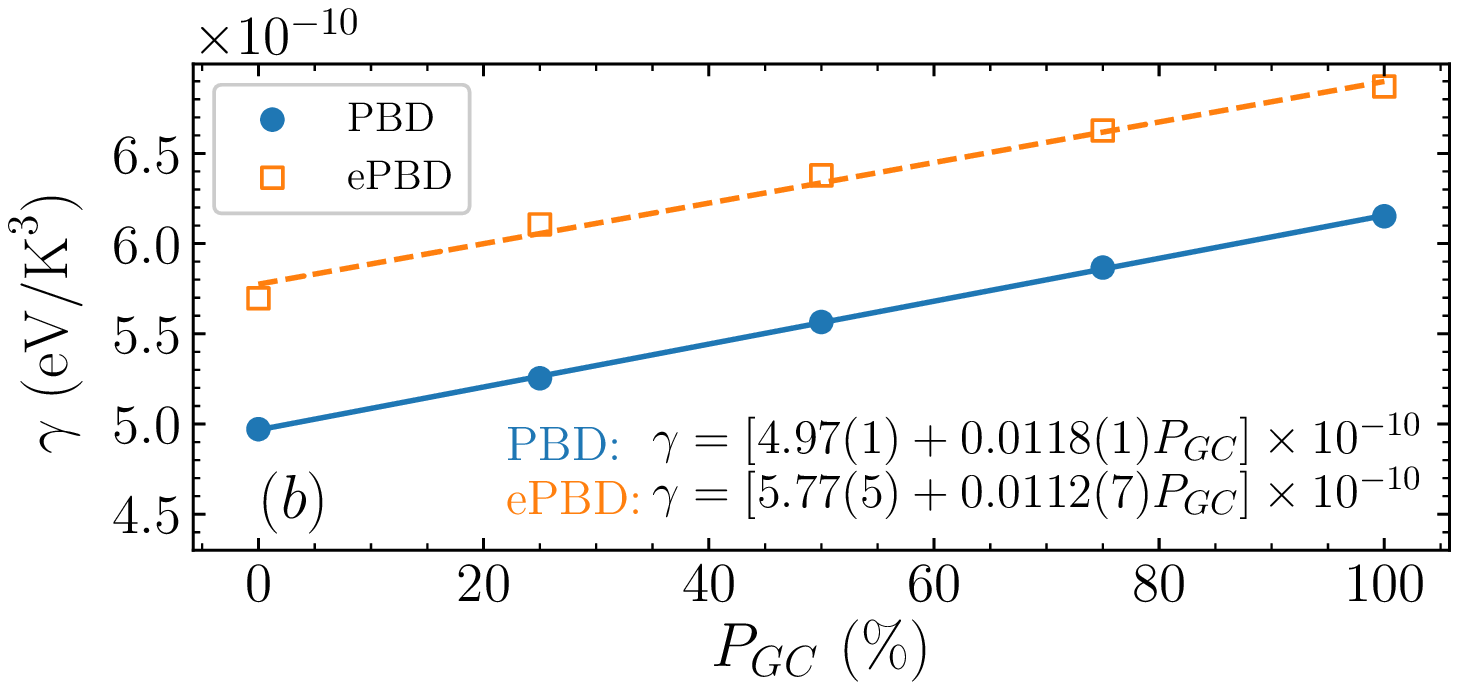}
    \caption{(a) Relationship between the energy density $E_N$ and the temperature $T$ of the ePBD model, for different
percentages $P_{GC}$ of GC base pairs in the DNA sequence (points). The data are fitted quite well with Eq.~\eqref{eq:cubicFit},
as shown by the curves in the plot. Results obtained for the PBD model are similar.
(b) Variation of the fitting parameter $\gamma$ in Eq.~\eqref{eq:cubicFit} with the GC percentage $P_{GC}$ of the DNA sequence,
showing a linear increase as $P_{GC}$ grows for both the PBD (blue circles) and the ePBD (empty orange squares) models.
Lines represent linear fits of the corresponding data, whose equations are also shown in the figure.
The values in the parentheses indicate the error of the computed fitting parameters, with for example $4.97(1)$
denoting $4.97 \pm 0.01$.}
    \label{fig:energyTemperature}
\end{figure}

Figure \ref{fig:energyTemperature}(a) shows the energy-temperature relation for the ePBD model, when DNA sequences
with various AT/GC composition (quantified by the percentage of GC base pairs, $P_{GC}$) are considered.
More specifically, results for a homogeneous DNA sequence consisted solely by AT ($P_{GC}=0\%$, blue circles)
or GC  ($P_{GC}=100\%$, purple squares) are presented, along with data for the heterogeneous case with
$P_{GC}=50\%$ (green triangles). Similar data for $P_{GC}=25\%$ and $P_{GC}=75\%$ have been also computed
(not shown in Fig. \ref{fig:energyTemperature}(a) for clarity). For all these cases, averaging was obtained
over 100 different realizations of DNA sequences with $N=1000$ base pairs each.
For the homogeneous cases 100 different initial conditions were created, while in the case of heterogeneous
DNA sequences with fixed $P_{GC}$, 100 different random arrangements of the AT and GC base pairs were considered
with random initial conditions. All these cases were integrated for 10 ns to allow the system's thermalization,
and then the temperature was recorded every picosecond for a further nanosecond. Averaging over all these numerical
results yields the final data points as those shown in Fig.~\ref{fig:energyTemperature}(a), where the computed standard
deviations give the presented error-bars. Results for the PBD model are very similar to those shown in Fig.~\ref{fig:energyTemperature}(a).
%%%%%%%%%%%%%%%%%%%%%%
%%%%%%%%%%%%%%%%%%%%%%

At low temperatures we see in Fig.~\ref{fig:energyTemperature}(a) a linear relationship between the energy density $E_N$
and the temperature $T$ of the form $E_N = k_B T$, as expected. As the temperature increases, a nonlinear dependence appears and
the addition of a simple cubic term provides a close fit to the data [see curves in Fig.~1(a)]. We used the fitting equation
\begin{equation}
    \label{eq:cubicFit}
    E_N = k_BT + \gamma T^3,
\end{equation}
with  $\gamma$ being a fitted constant. Applying a least-squares fitting algorithm \cite{More1977}, we find Eq.~\eqref{eq:cubicFit}
to approximate very well the numerical data for both models, at all $P_{GC}$ percentages.
The resulted values of the fitting parameter $\gamma$ are shown in Fig.~\ref{fig:energyTemperature}(b). For both systems
the obtained values of the coefficient $\gamma$ are of the order of 10$^{-10}$ eV/K$^3$ and are very well represented
by linear functions of the percentage $P_{GC}$, as shown by the straight lines in Fig.~\ref{fig:energyTemperature}(b).
The parameter $\gamma$ of the ePBD model is shifted to higher values than that of the PBD system, indicating that
the ePBD energy is slightly above the corresponding PBD energy for larger temperatures.
This suggests that the lower average stacking energy of the ePBD model (see figure 2 of
Ref.~\cite{Alexandrov2009ePBD}) results in a slightly higher overall energy as compared to the PBD model at the same temperature.

The calculated energy-temperature relations will be used in the following Sections in order to obtain results
corresponding to fixed temperatures, through our microcanonical numerical computations. In particular, to simulate the PBD
or ePBD system at a desired temperature, we determine its conserved energy density through the respective $E_N-T$
relation and then follow the numerical integration procedure mentioned above.

%=========================
\section{Bubble Opening Thresholds} % (fold)
\label{sec:bubble_opening_thresholds}

In order to effectively investigate statistical properties of bubble openings, we first define a threshold
for considering a base pair to be separated. In various studies, the thresholds used for this purpose range
from around 0.5 \AA\ up to 5 \AA\ or more, depending on the particular application (see
e.g.~\cite{Choi2004,Kalosakas2004,Choi2008,Alexandrov2009,Alexandrov2010,Apostolaki2011,faloPRE12,huangJBE}).
Here we choose a threshold that is able to reproduce the known melting behavior of DNA molecules in the PBD
model \cite{Kalosakas2009}, taking into account that by definition at the melting transition 50\% of base pairs are separated.
Thus the requirement is for our threshold to mark 50\% of base pairs open at the melting temperature, for sequences of varied AT and GC base pair compositions.

Actually the characteristic length of the intra-base pair Morse potential in Eq.~(\ref{eq:Morse}), $1/a_{GC}$ and
$1/a_{AT}$ for GC and AT base pairs respectively, provides a physical choice that turns out to fulfil our requirements
on such a threshold. It is important to note here that we are using a different opening threshold for AT and GC base pairs.
The use of a common threshold is not so consistent with the requirement of 50\% open base pairs at melting.
On the other hand, it is reasonable to consider different thresholds for the opening of GC and AT base pairs due to the variation
of the parameters describing the corresponding on-site Morse potential.

\begin{figure}[t]
    \centering
    \includegraphics[width=0.48\textwidth]{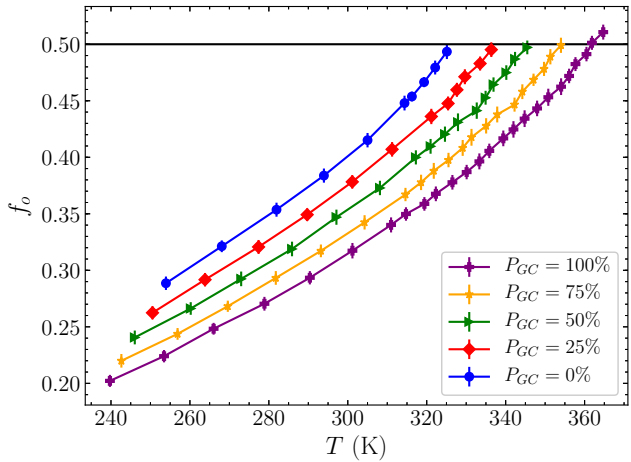}
    \caption{The fraction $f_o$ of open base pairs as a function of temperature $T$ in the PBD model, for chains of various
$P_{GC}$ percentages, stopping at the melting temperature in each case (points). Data are line-connected to guide the eye.
The thresholds used here for considering a base pair as open are $y_{GC}^{thr}=0.15$ \AA\ and $y_{AT}^{thr}=0.24$ \AA\
(see text for details). The horizontal, solid black line indicates the value $f_o=0.5$, i.e.~50\% of base pairs are open.}
    \label{fig:thresholdsPBD}
\end{figure}

In Fig.~\ref{fig:thresholdsPBD}, we see for the PBD model the increase in the fraction of open base pairs $f_o$ with temperature $T$,
up to the melting temperature defined by $T_m^{PBD} = 325+0.4P_{GC}$ \cite{Kalosakas2009}, for the proposed thresholds
of $y_{GC}^{thr}=1/a_{GC} = 0.15$ \AA\ and $y_{AT}^{thr} = 1/a_{AT} = 0.24$ \AA\ for GC and AT base pairs respectively.
Each point is the averaged fraction of open base pairs over 100 different realizations of DNA sequences with $N=1000$ base pairs, apart from the last five points (closest to the melting transition) shown in each case, where 200 runs were used
in order to have better statistics in the region of interest.
At the melting point (corresponding to the high temperature end of the presented data), almost exactly 50\% of the base pairs
are open (the value $f_o=0.5$ is indicated by the horizontal, solid line in Fig.~\ref{fig:thresholdsPBD}).
These results indicate that the proposed thresholds can be efficiently used as appropriate measures for considering
base pairs to be open. It is worth noting that since in the PBD model a scaling factor of $1/\sqrt{2}$ is applied
to the stretchings $y_n$ \cite{Peyrard1993,Kalosakas2004}, the actual relative displacements of complementary bases
represented by these thresholds are $0.21$ \AA\ and $0.34$ \AA\ for GC and AT base pairs respectively.
%%%%%%%%%%%%%%%%%%%%%%
%%%%%%%%%%%%%%%%%%%%%%

\begin{figure}[t]
    \centering
    \includegraphics[width=0.48\textwidth]{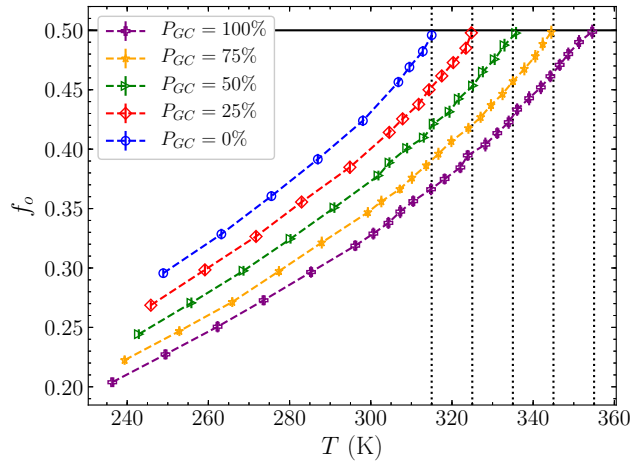}
    \caption{The fraction $f_o$ of open base pairs at a given temperature $T$ for the ePBD model with various $P_{GC}$
percentages (points), stopping when $f_o=0.5$ (horizontal solid line) in each case. Data are line-connected to guide the eye.
 The thresholds for considering a base pair as open are as in Fig.~\ref{fig:thresholdsPBD}.
 Vertical dotted lines indicate the estimated melting temperatures $T_m$ for each $P_{GC}$ value.}
    \label{fig:thresholdsePBD}
\end{figure}

Noting that the Morse potential, Eq.~\eqref{eq:Morse}, governing the intra-base pair interactions remains unchanged in the ePBD model, we can implement the same threshold values as in the PBD system for defining the opening of a base pair in the ePBD case.
Then, repeating in Fig.~\ref{fig:thresholdsePBD} for the ePBD model similar calculations to the ones presented in
Fig.~\ref{fig:thresholdsPBD}, we are able to obtain the melting temperatures of the ePBD system as the temperature values
at which the fraction of open base pairs is $f_o=0.5$, without going through the detailed procedure implemented for the PBD
model in Ref.~\cite{Kalosakas2009}. This approach allows us to estimate the melting temperature $T_m^{ePBD}$ of the ePBD model
for various $P_{GC}$ levels. The corresponding $T_m^{ePBD}$ values are indicated by the vertical lines in
Fig.~\ref{fig:thresholdsePBD} and they are accurately obtained (in K) by the relation
\begin{equation}
    \label{eq:TM}
    T_m^{ePBD} = 315 + 0.4P_{GC},
\end{equation}
which retains the experimentally observed linear relationship between the melting temperature and the GC percentage,
exhibiting a slope in quantitative agreement with the measured value \cite{exper}.
%%%%%%%%%%%%%%%%%%%%%%
%%%%%%%%%%%%%%%%%%%%%%

%=========================
\section{Bubble Lifetime Distributions} % (fold)
\label{sec:bubble_lifetime_distributions}

Based on the base pair opening thresholds determined in Sect.~\ref{sec:bubble_opening_thresholds}, we are now investigate
in detail the statistical properties of bubbles in DNA at $T=310$ K. By performing constant energy molecular dynamics (MD)
simulations, we track the creation and destruction of bubbles, and record their lifetimes. Our microcanonical simulations
differ from previous studies of bubble lifetimes using Langevin MD \cite{Alexandrov2006,Alexandrov2009, Alexandrov2010}.
The microcanonical ensemble probes the
inherent characteristic times of the model, in contrast to the Langevin dynamics which introduces artificial
time scales through the arbitrary damping coefficient. On the other hand, there are of course benefits
to considering the fluctuations provided by the random forces in Langevin dynamics, in order to better mimic a heat bath
at finite temperatures and assess \textit{relative} timescales of different fluctuations. However, even in this case it is not known whether the white noise of the stochastic term
in Langevin simulations realistically describes the biological environment, or a coloured noise with
specific characteristics is more appropriate to describe the interactions of DNA with its surroundings.
Therefore we have preferred as a first step to investigate the inherent dynamics of the system using microcanonical
simulations, while its temperature is effectively represented through its energy density $E_N$ as it is described in
Section~\ref{sec:dna_models} (see Eq.~\ref{eq:cubicFit}). It is not obvious what effect Langevin dynamics would have
on the calculated distributions and how the arbitrary friction constant affects the lifetime distributions.
Such an investigation is left for a future work.
Obtaining statistically sound bubble lifetime distributions for different bubble lengths is a computationally nontrivial task,
due to both the large amounts of data required and the complexity of the problem of identifying and tracking bubbles accurately.

To clarify the method we used to obtain bubble lifetime distributions, the outline of the implemented algorithm
for the production of the distributions for an individual realization is as follows:
\begin{enumerate}
    \item Perform MD simulations to create records of open/closed information for each base pair in the DNA sequence at each time step.
    \item At each time step, look along the sequence and record the length of any occurring bubble, attributed at the corresponding starting site.
    \item Check each bubble (site and length) against the previous time step.
    \begin{itemize}
        \item If a bubble occurs somewhere that there was no bubble previously, begin a record of that bubble -- a tuple of (length, lifetime).
        \item If a bubble survives identically, increment the lifetime of that bubble by one time step.
        \item If a bubble changes length, close the record of that bubble, and start a new record at that site with the new length.
        \item If no bubble is present somewhere that it was existing a bubble previously, close that record.
    \end{itemize}
    \item At the end of the simulation, record the list of (length, lifetime) tuples at each site.
\end{enumerate}

We note that our simple bubble-tracking criteria are fairly strict, in that small fluctuations at the end of a given
bubble, due to transient openings or closings of base pairs at its ends, would result in starting new bubbles from the beginning.
While this choice provides a straightforward and efficient method of lifetime calculation, other more flexible criteria
allowing small fluctuations of bubbles when computing their lifetimes may be relevant too. Further, other algorithms
measuring the lifetime of bubbles with size larger than a particular length may also be of interest, in particular for exploring
bubble dynamics in biologically functional DNA sequences.

Data from many runs can be combined to create statistically meaningful distributions.
In our investigation, for each case studied (different AT/GC content) we have in generally used 1000 different realizations
of $N=100$ base pair long DNA sequences, integrated until 10 ns for thermalization, and then data recorded
every 0.01 ps for the next 1 ns.
In order to establish the accuracy of the implementation of the used algorithm, in the absence of existing results
for bubble lifetimes by microcanonical simulations, tests were performed against artificially created data sets
with known distributions, and the full analysis as outlined above was performed on these data sets. The code exactly
reproduced the known distributions, providing an assurance about the reliability of the results presented below.

%%%%%%%%%%%%%%%%%%%%%%
\begin{figure}[tb]
    \centering
    \includegraphics[width=0.48\textwidth]{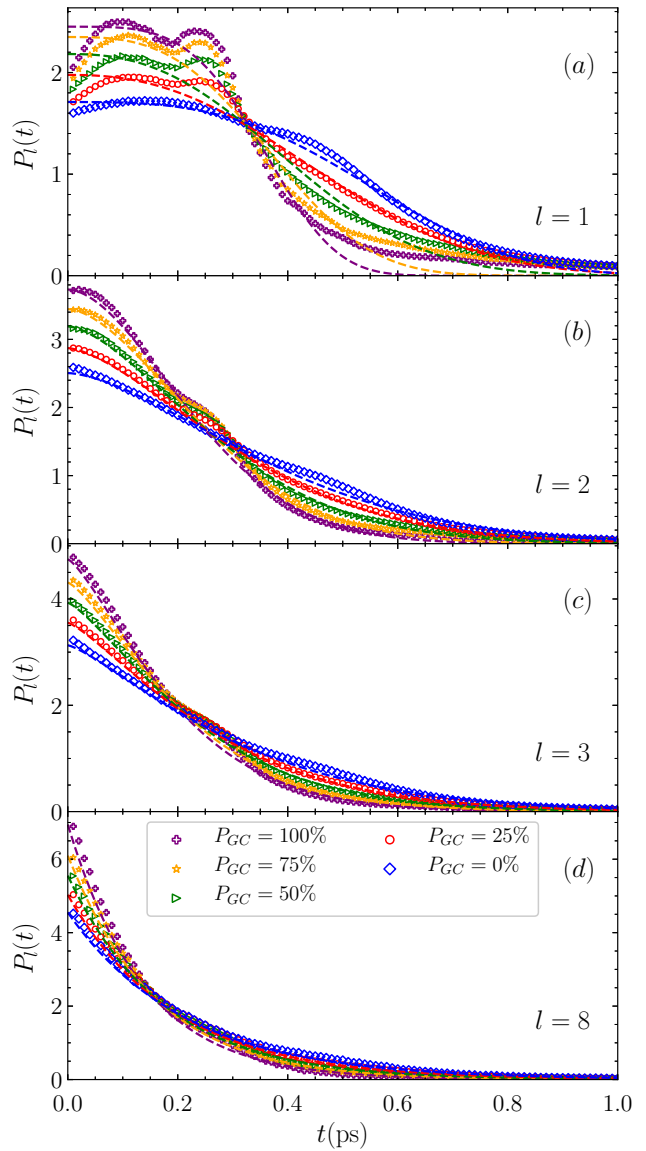}
    \caption{Bubble lifetime distributions $P_l(t)$ for different GC percentages $P_{GC}$, in the case of bubble length (a) $l=1$, (b) $l=2$, (c) $l=3$, and (d) $l=8$ (points). Results are shown for the ePBD model, with similar distributions being obtained for the PBD system. The dashed curves correspond to fits of the data with the stretched exponential function of Eq.~\eqref{eq:stretchedExpLife}.}
    \label{fig:lifetimesDist}
\end{figure}
%%%%%%%%%%%%%%%%%%%%%%

Based on the data obtained with this approach we first examine the effect of the AT/GC composition of DNA molecules
on the bubble lifetime distributions $P_l(t)$, for different bubble lengths $l$. Representative distributions
for several bubble lengths and GC percentages are shown with points in Fig.~\ref{fig:lifetimesDist}, illustrating an
approximately exponential profile with the exception of single base pair openings for $l=1$ [Fig.~\ref{fig:lifetimesDist}(a)].
Data for nine different $P_{GC}$ percentages have been obtained, but for clarity we do not present
all of them in Fig.~\ref{fig:lifetimesDist}. Further, lifetime distributions $P_l(t)$ for bubble lengths $l=1,2,\ldots,10$
have been calculated, but the cases $l \geq 4$ are very similar and thus only the $P_{l=8}(t)$ is shown in Fig.~\ref{fig:lifetimesDist}(d).
Only results for the ePBD model are shown in Fig.~\ref{fig:lifetimesDist}, as on this scale the difference between
the PBD and ePBD data is very small.

We see from Fig.~\ref{fig:lifetimesDist}(a) that in the case of bubbles with $l=1$ a two-peaked profile is present,
with the height of these two peaks depending on the GC content. Apart from the case with $P_{GC} = 0\%$, the two peaks are
visible around $t=0.1$ ps and $t=0.25$ ps. In the case of homogeneous AT sequences ($P_{GC} = 0\%$) the two peaks are
very broad, located around $t=0.15-0.20$ ps for the first one and around $t=0.4-0.5$ ps for the second.
The positions of these peaks are not related to the periods of the $q=\pi$ vibrational normal modes of GC
or AT base pairs (at around 0.4 ps and 0.8 ps, respectively, which are further increased due to thermal softening at $T=300$ K
\cite{Voulgarakis2004}). The particular complex structure of these distributions may arise from
the interplay of the characteristic times of single base pair bubbles and the transient $l=1$ base pair
openings of either increasing or decreasing in size larger bubbles during their opening and closing respectively.
As evidenced in Figs.~\ref{fig:lifetimesDist}(b) and \ref{fig:lifetimesDist}(c) some peaks can be still distinguished
in the cases of $l=2$ and $l=3$, but they become less prominent a $l$ increases. For longer bubbles a smoothing out
of these peaks is observed, as seen for example in Fig.~\ref{fig:lifetimesDist}(d).

The bubble lifetime distributions $P_l(t)$ can be fitted quite well with a stretched exponential function,
\begin{equation}
    \label{eq:stretchedExpLife}
    P_l(t) = A\exp\left(-\left(t/\tau\right)^{\beta}\right),
\end{equation}
for all cases apart from $l=1$, where the stretched exponential parameters $\beta$ and $\tau$ depend on $l$ and $P_{GC}$.
These fits are shown for all cases in Fig.~\ref{fig:lifetimesDist} by dashed curves. When the corresponding curve is not
visible, it is covered by the overlying data points. From Fig.~\ref{fig:lifetimesDist}(a) we see that the stretched
exponential distribution, Eq.~\eqref{eq:stretchedExpLife} does not capture of course the somewhat complex double-peaked
profile for $l=1$, but still provides a rough approximation of the overall behavior.
For $l\geq 2$  however, the stretched exponential of Eq.~\eqref{eq:stretchedExpLife} describes the numerical data much more accurately
[see Figs.~\ref{fig:lifetimesDist}(b)-(d)], and can be used to meaningfully approximate the bubble lifetime distributions.

%%%%%%%%%%%%%%%%%%%%%%
\begin{figure}
    \centering
\includegraphics[width=0.48\textwidth]{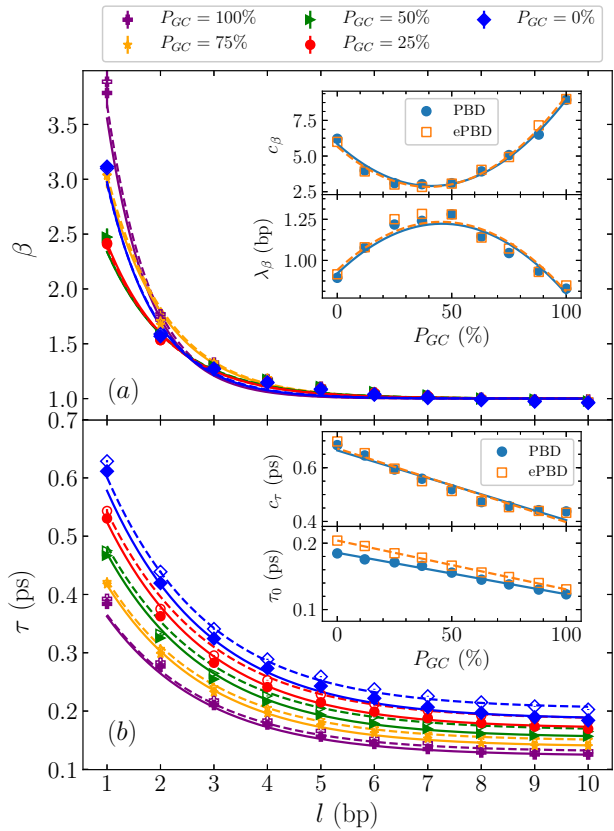}
    \caption{The fitting parameters of the bubble lifetime distributions $P_l(t)$ of Eq.~\eqref{eq:stretchedExpLife} for
different $P_{GC}$ percentages (points shown by different colors): (a) the stretched exponent $\beta$ and
(b) the characteristic time $\tau$, with respect to the bubble length $l$ [in number of pase pairs (bp)].
The independencies on $l$ of both of these parameters are fitted with the exponentially decaying functions, Eqs. \eqref{eq:xifit} and \eqref{eq:taufit}.
In (a) and (b) filled (empty) symbols indicate results for the  PBD (ePBD) model and solid (dashed) curves show fits with
Eq.~\eqref{eq:xifit} and Eq.~\eqref{eq:taufit} respectively.
The insets in (a) present the dependence of the parameters $c_\beta$ and $\lambda_\beta$ of Eq.~\eqref{eq:xifit}
on the GC content of the sequence (quantified by the $P_{GC}$ value) for the PBD (blue circles) and the
ePBD (empty orange squares) models. Solid and dashed curves represent appropriate fits of the corresponding PBD and ePBD
data with quadratic functions. These parameters are almost identical for the two models.
The insets in (b) are similar to the ones in (a) but for the $c_{\tau}$ and $\tau_0$ parameters of Eq.~\eqref{eq:taufit}.
    }
    \label{fig:lifetimeParams}
\end{figure}
%%%%%%%%%%%%%%%%%%%%%%

The values of the numerically obtained fitting parameters $\tau$ and $\beta$ of Eq.~\eqref{eq:stretchedExpLife}
are shown in Fig.~\ref{fig:lifetimeParams}.
It is apparent, already by inspection of Fig.~\ref{fig:lifetimesDist}, and, more precisely, from the behavior of the
stretched exponent $\beta$ in Fig.~\ref{fig:lifetimeParams}(a) (which practically becomes $\beta=1$ for larger $l$ values,
irrespective of the GC percentage), that the $P_l(t)$ distributions of Eq.~\eqref{eq:stretchedExpLife} become more closely
exponential as the bubble length $l$ increases.
In fact, because $\beta>1$ for short bubbles, the corresponding distributions show a compressed exponential
behavior meaning that there exist more short-lived bubbles (with lifetimes smaller than $\tau$) and less long-lived bubbles
(with lifetimes larger than $\tau$) in these cases as compared to a purely exponential distribution. This leads to average
lifetimes which are shorter than the characteristic lifetimes $\tau$, as we will see below (cf. Fig.~\ref{fig:lifetimesMean}).
This functional dependence reflects the extreme rarity of large long-lived
bubbles in arbitrary DNA sequences. It is worth noting that as we see from the data of Fig.~\ref{fig:lifetimeParams}(a)
the values of the $\beta$ exponent decay exponentially and they are practically the same for the PBD
(filled symbols and solid lines) and the ePBD (empty symbols and dashed lines) models, at any GC content.

From the results of Fig.~\ref{fig:lifetimeParams}(b) we see that the characteristic time $\tau$ of
Eq.~\eqref{eq:stretchedExpLife} also decreases exponentially with bubble length, up to an asymptotic value
dependent on the GC content. The PBD and ePBD models give a little different values for $\tau$, with  slightly longer characteristic times observed always in the ePBD model, while the difference is more
noticeable as the AT content of the sequence increases.
This, taking also into account that the exponent $\beta$ is practically the same for both models, suggests that the ePBD model
exhibits typically longer-lived bubbles than the PBD model.

The variation of both parameters $\beta$ and $\tau$ of Eq.~\eqref{eq:stretchedExpLife} with the bubble length $l$
can be fitted with simple exponentials, of the form
\begin{align}
    \label{eq:xifit}\beta &= c_\beta\exp(-l/\lambda_\beta)+1,\\
    \label{eq:taufit}\tau &= c_\tau\exp(-l/\lambda_\tau)+\tau_0.
\end{align}
As already mentioned, the $\beta$ values are almost indistinguishable for the PBD and ePBD models.
This is also reflected by the fact that the computed $c_\beta$ and $\lambda_\beta$ values of Eq.~\eqref{eq:xifit} for
various GC contents are practically identical for both DNA models, as shown in the insets of Fig.~\ref{fig:lifetimeParams}(a).
Thus, the dependence of $c_\beta$ and $\lambda_\beta$ on $P_{GC}$ can be very well approximated by the same quadratic
functions for the PBD and the ePBD models, and the corresponding fitted equations are
$c_\beta = 0.0017(1)(P_{GC})^2-0.14(1)P_{GC}+5.9(2)$ and $\lambda_\beta = -0.00015(2)(P_{GC})^2 + 0.014(2)P_{GC} + 0.93(4)$.
Thus, for both DNA models the bubble lifetime distributions $P_l(t)$, Eq.~\eqref{eq:stretchedExpLife}, approach simple
exponential functions for larger bubble lengths $l$ at the same way, as the exponent $\beta$ tends towards 1
identically in both cases.

We also find that the values of $c_\tau$ and $\lambda_\tau$ in Eq.~\eqref{eq:taufit} are similar for the two models.
As demonstrated in the upper inset of Fig.~\ref{fig:lifetimeParams}(b), $c_\tau$ varies linearly with the GC percentage,
fitted by $c_\tau = 0.66(1)-0.0026(2)P_{GC}$ for both models, while $\lambda_\tau = 1.9$ bp is constant across all
compositions for both PBD and ePBD cases. On the other hand, as we see in the lower inset of Fig.~\ref{fig:lifetimeParams}(b),
the asymptotic value $\tau_0$ in Eq.~\eqref{eq:taufit} shows a linear decrease with $P_{GC}$ for both systems,
while it is always slightly larger for the ePBD model. In particular, this linear dependence can be fitted by
$\tau_0 = 0.19(1)-0.0006(2)P_{GC}$, for the PBD model and $\tau_0 = 0.20(1) - 0.0007(2)P_{GC}$ for the ePBD model.

These results show that the difference between the two models in random averages is only evident in the
linear shift of the asymptotic value $\tau_0$ of $\tau$ in Eq.~\eqref{eq:taufit}, with the shape of the distributions $P_l(t)$,
Eq.~\eqref{eq:stretchedExpLife}, otherwise being very similar. In our computations the normalization constant $A$ in
Eq.~\eqref{eq:stretchedExpLife} was considered as a free fitted parameter. The numerically obtained $A$ values quite
accurately reproduce the normalization condition $\int_0^{\infty} P_l(t) dt =1$, as this property was recovered with an overall
discrepancy of around 5\%.

%%%%%%%%%%%%%%%%%%%%%%
\begin{figure}[t]
    \centering
    \includegraphics[width=0.48\textwidth]{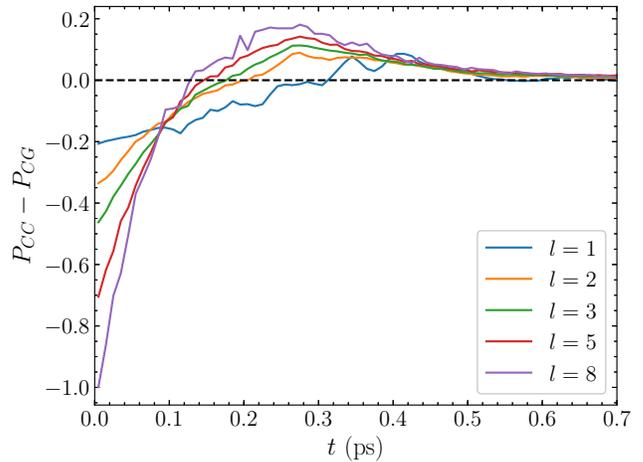}

    \caption{he difference $P_{CC}-P_{CG}$ in the ePBD model between the bubble lifetime probability
distributions for DNA sequences containing only C bases along one strand (and G bases along the complementary one),
and sequences of alternating CG bases along each strand. The positive difference at longer lifetimes indicates the tendency
of longer-lived bubbles to be formed in the homogeneous sequence as compared to the CG periodic repeats.
Each line corresponds to a different bubble length, as shown in the legend.}
    \label{fig:comparison_cccg}
\end{figure}
%%%%%%%%%%%%%%%%%%%%%%

he advantage of the ePBD model is that it can accurately predict the thermal openings and denaturation
temperatures of homogeneous and periodic DNA sequences exhibiting unusual melting transitions, where the original
PBD model makes no distinction \cite{Alexandrov2009ePBD}. A characteristic example is provided by the
homogeneous (C)$_{36}$ and the periodic (CG)$_{18}$ oligonucleotides, where their melting temperatures differ by
more than 20 degrees (74 C and 96 C, respectively \cite{Alexandrov2009ePBD}). The PBD model cannot distinguish
these two sequences. 
On the contrary the ePBD model can successfully describe their different melting behavior through the different stacking constants $K_{CC}$ in the
former sequence and $K_{CG}$, $K_{GC}$ in the latter one. 
Even though the averaged results on random sequences presented in this work show small quantitative differences between the PBD and ePBD models, when such specific DNA segments are considered then the ePBD model provides more accurate calculations of the bubble distributions.
To explicitly demonstrate this, we have calculated the bubble lifetime distributions for the homogeneous (C)$_{36}$
and the periodic (CG)$_{18}$ DNA segments of 36 base pairs using periodic boundary conditions in both cases.
For simplicity we refer to these sequences as CC and CG respectively, and the corresponding bubble lifetime distributions are denoted as $P_{CC}$ and $P_{CG}$.
The PBD model obviously gives identical distributions $P_{CC}$ and $P_{CG}$.
These are very similar to the $P_{CG}$ obtained by the ePBD model, due to the almost identical value of $K_{GC}$ with the
parameter $k$ of PBD and the relatively nearby value of $K_{CG}$ (see Section~\ref{sec:dna_models}).
However the ePBD model results in systematic differences in the $P_{CC}$ distributions due the much smaller value
of $K_{CC}$. This is shown in Fig.~\ref{fig:comparison_cccg} through the difference $P_{CC}-P_{CG}$, computed using the average over 4000 realisations.
A systematic variation can be seen in these distributions for all bubble lengths examined here, as $P_{CC}$ is smaller
than $P_{CG}$ for relatively short-lived bubbles (indicating more such bubbles in the CG alternating sequences), while it is
the other way around for longer-lived bubbles (revealing more long-lived bubbles in the homogeneous CC segment, having
a lower stacking energy). 
At their largest, these differences are greater than 10\% of the distribution values at that point.
These results show the effect of the sequence-dependent stacking encoded in the ePBD model
on the bubble lifetimes of specific DNA stretches.

%%%%%%%%%%%%%%%%%%%%%%
\begin{figure}[t]
    \centering
    \includegraphics[width = 0.48\textwidth]{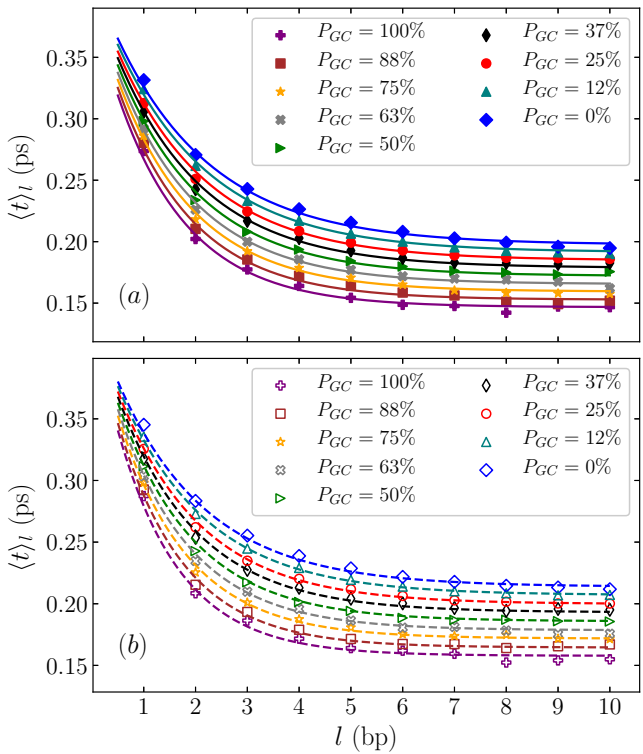}
    \caption{
  Mean bubble lifetimes $\langle t \rangle_l$ as a function of bubble length $l$ for (a) the PBD and (b) the ePBD model,
for different $P_{GC}$ percentages (points). Solid and dashed curves in, respectively (a) and (b) show fits of the data with Eq.~\eqref{eq:meanExponential}.}
    \label{fig:lifetimesMean}
\end{figure}
%%%%%%%%%%%%%%%%%%%%%%

We can numerically estimate the  mean bubble lifetime $\langle t \rangle_l$ according to
\begin{equation}
    \label{eq:numericalMean}
    \langle t \rangle_l = \sum_{i=1}^{M}t_i P_l(t_i)\delta t
\end{equation}
where $P_l(t_i)$ is the numerically estimated probability density of bin $i$ with width $\delta t$, and $t_i$ is the time
at the middle of that bin. As this sum is finite and based on the fact that $P_l(t)$ practically vanishes for
relatively large $t$, the $\langle t \rangle_l$ value in Eq.~\eqref{eq:numericalMean} is computed by considering
$M=500$ bins of width $\delta t = 0.01$ ps. The obtained results are presented in Figs.~\ref{fig:lifetimesMean}(a)
and \ref{fig:lifetimesMean}(b) for the PBD and ePBD model, respectively. We see that the mean bubble lifetime decreases
exponentially with bubble length $l$. A clear monotonic decrease in bubble lifetimes with increasing $P_{GC}$ values is also
evident at every bubble length. By comparing Figs.~\ref{fig:lifetimesMean}(a) and \ref{fig:lifetimesMean}(b) we see that
the ePBD model exhibits slightly higher average lifetimes, but nevertheless shows the same trend as the PBD model.
These longer average lifetimes in the ePBD model are consistent with results found in Ref.~\cite{Alexandrov2009ePBD}
showing that in general larger base pair displacements are observed in the ePBD than the PBD model (see figure 4
of that reference), taking also into account that the same opening thresholds are considered here in both cases.
%%%%%%%%%%%%%%%%%%%%%%
\begin{figure}[t]
    \centering
    \includegraphics[width = 0.48\textwidth]{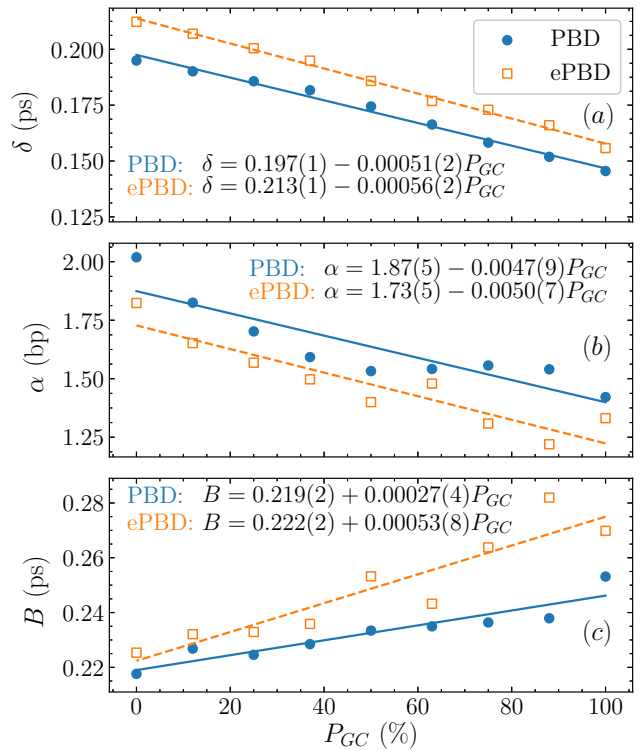}
    \caption{ Dependence of (a) the asymptotic value $\delta$, (b) the characteristic length $\alpha$, and
(c) the prefactor $B$, of the fitting of the $\langle t \rangle_l$ numerical data in Fig.~\ref{fig:lifetimesMean}
with Eq.~\eqref{eq:meanExponential} on the GC content of the DNA sequence.
  Blue circles show results for the PBD model and empty orange squares for the ePBD model. Lines show linear fits of the presented data
  with the equations reported at the corresponding panel.}
    \label{fig:lifetimesMean_params}
\end{figure}
%%%%%%%%%%%%%%%%%%%%%%

The dependence of the mean bubble lifetime $\langle t \rangle_l$ on the bubble's length $l$ for both PBD and ePBD models is
accurately fitted through a simple exponential decay of the form \begin{equation}
    \label{eq:meanExponential}
    \langle t \rangle_l = B\exp\left(-l/\alpha\right) + \delta,
\end{equation}
as shows the good description of  the data points in Figs.~\ref{fig:lifetimesMean}(a) and \ref{fig:lifetimesMean}(b) by the solid
and dashed curves respectively.
The $P_{GC}$ dependence of the three free parameters of Eq.~\eqref{eq:meanExponential}, namely the asymptotic value $\delta$
[see Fig.~\ref{fig:lifetimesMean_params}(a)], the characteristic length $\alpha$ [see Fig.~\ref{fig:lifetimesMean_params}(b)],
and the prefactor $B$ [see Fig.~\ref{fig:lifetimesMean_params}(c)] is reasonably approximated by linear fits. These are shown
by solid blue and dashed orange straight lines in Figs.~\ref{fig:lifetimesMean_params}(a)-(c) along with the corresponding
PBD (blue circles) and ePBD (empty orange squares) data.

Closing this section, we note that the characteristic times of the bubble lifetimes calculated here are of the order of
$\sim$ 10$^{-1}$ ps. This time scale coincides with the faster relaxation time (between at least two distinct relaxation
processes appeared in the range from $10^{-2}$ up to $3 \times 10^3$ ps) observed in the time-dependent autocorrelation
functions of base pair fluctuations in the PBD model for homogeneous (purely AT or GC) DNA sequences
(see figures 1 and 2 of Ref.~\cite{Kalosakas2006}). In that work, local fluctuations of base pair openings were considered
(corresponding to $l=1$), while microcanonical MD was also used.

%=========================
\section{Bubble Length Distributions}
\label{sec:bubble_length_distributions}

Let us now discuss  the distribution of bubble lengths based on our MD simulations.  Investigations of such distributions
and their dependence on GC content have already been performed using Monte Carlo simulations, at physiological
temperature~\cite{Ares2007}, as well as in the temperature range 270-350 K \cite{Kalosakas2009}. A uniform threshold of
$y_n=1.5$ \AA\ was used for both types of base pairs in those studies. Here we use extensive MD calculations and the base pair specific
thresholds defined in Sect.~\ref{sec:bubble_opening_thresholds} to examine these distributions, at a fixed temperature of $T=310$ K.
For this purpose we perform  simulations for DNA sequences of $N=1000$ base pairs, considering  8000 different, random realizations.
Each case is  again integrated for 10 ns to ensure thermalization, and then bubble length data are recorded every 0.1 ns
for a further 10 ns. These conditions ensure a quite rich statistics, which is necessary for the accuracy of the tails of the distributions
for bubble lengths of the order of tens of base pairs.

%%%%%%%%%%%%%%%%%%%%%%
\begin{figure}[t]
    \centering
    \includegraphics[width=0.48\textwidth]{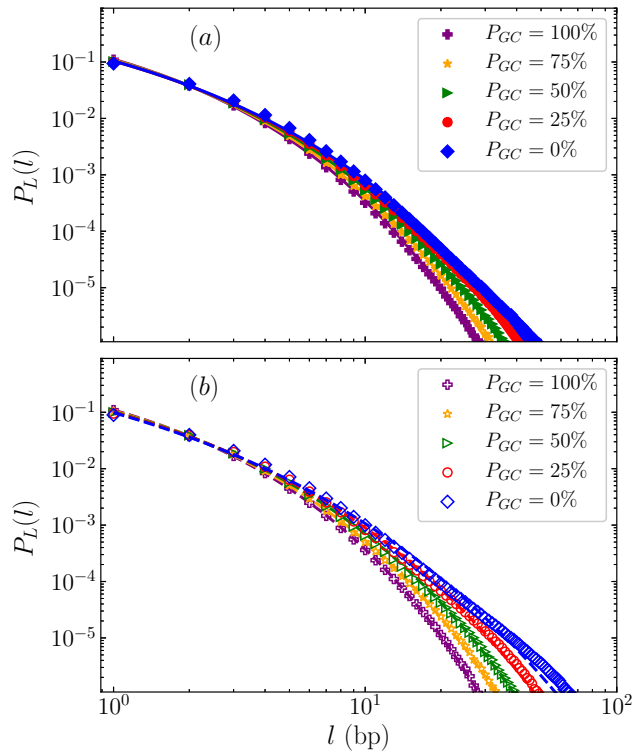}
    \caption{The probability $P_L(l)$ for the appearance of bubbles of length $l$ in double stranded DNA at physiological
temperature $T=310$ K, for various $P_{GC}$ percentages, plotted in log-log scale for (a) the PBD model and
(b) the ePBD model (points). Solid curves in (a) and dashed curves in (b) depict fits of the data with Eq.~\eqref{eq:stretchFitLength}.}
    \label{fig:bubbleLengths}
\end{figure}
%%%%%%%%%%%%%%%%%%%%%%

Corresponding results are shown in Fig.~\ref{fig:bubbleLengths}. Distributions of bubble lengths $P_L(l)$ for different
GC percentages at $T=310$ K are presented in Fig.~\ref{fig:bubbleLengths}(a) for the PBD and in Fig.~\ref{fig:bubbleLengths}(b)
for the ePBD model. Similar data have been obtained for four more $P_{GC}$ cases, in between of those values
depicted in Fig.~\ref{fig:bubbleLengths}, not shown here for clarity.
From the profiles of these distributions we see that for short bubble lengths ($l \lesssim 4$) the probabilities are relatively unaffected
by the base pair content, as practically the $P_L(l)$ results coincide for all $P_{GC}$ values. However for longer bubbles,
the GC content of the DNA sequence  plays a significant role on the bubble length probabilities as different $P_L(l)$ values
are observed at different $P_{GC}$ levels. In particular, in this case AT-rich strands exhibit noticeably more large bubbles
than GC-rich sequences in both models, as expected. While this behavior, at larger bubble lengths, is in accordance to what
has been previously observed for the PBD model using uniform thresholds~\cite{Kalosakas2009}, the shorter length insensitivity
of the distributions on the base pair content of the sequence shown in Fig.~\ref{fig:bubbleLengths} is unique for the base pair
specific thresholds considered here.

The ePBD model [Fig.~\ref{fig:bubbleLengths}(b)] favors the appearance of more bubbles at all lengths and GC contents
with respect to the PBD system [Fig.~\ref{fig:bubbleLengths}(a)], in line with the overall lower melting temperatures
exhibited by this model, although for the case of pure GC sequences ($P_{GC}=100\%$) the two models give quite similar
results. The differences between the PBD and the ePBD model become more pronounced as more AT base pairs are added
to the sequence, with the pure AT sequences ($P_{GC}=0\%$) showing a distinctive feature in the tail of the probability
distribution for longer bubbles ($l>30$ bp) in the ePBD case.

%%%%%%%%%%%%%%%%%%%%%%
\begin{figure}[t]
    \centering
    \includegraphics[width=0.48\textwidth]{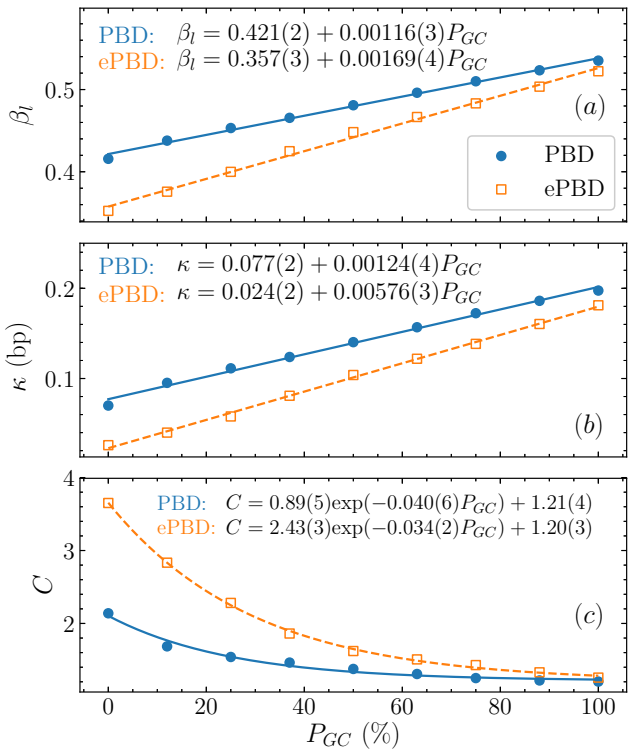}
    \caption{ Dependence on the GC content of the DNA sequence of (a) the stretched exponent $\beta_l$,
(b) the characteristic length $\kappa$ and (c) the prefactor $C$ of the fit of the $P_L(l)$ distributions shown
in Fig.~\ref{fig:bubbleLengths} with Eq.~\eqref{eq:stretchFitLength},
for the PBD (blue circles) and the ePBD (empty orange squares) models. Fits of the presented data with a straight line in (a) and (b)
and an exponential function in (c), are shown, and the corresponding fitting equations are reported in each panel.}
    \label{fig:bubbleLengths_params}
\end{figure}
%%%%%%%%%%%%%%%%%%%%%%

The numerically computed distributions of Figs.~\ref{fig:bubbleLengths}(a) and \ref{fig:bubbleLengths}(b) can be suitably
fitted with a stretched exponential function,
\begin{equation}
    \label{eq:stretchFitLength}
    P_L(l) = C\exp\left(-\left(l/\kappa\right)^{\beta_l}\right),
\end{equation}
as can be seen by the solid and dashed curves, respectively.
In a previous work it has been found that this functional form provided an accurate fitting of the bubble length distributions
of PBD, equally well with a power-law modified exponential~\cite{Ares2007}. Here, however, the latter function
cannot describe satisfactorily the tails of the distribution for the AT-rich ePBD case, in contrast to Eq.~\eqref{eq:stretchFitLength}.

The numerical values of the free parameters of the fitting with Eq.~\eqref{eq:stretchFitLength}, namely the characteristic length
$\kappa$, the stretched exponent $\beta_l$, and the preexponential coefficient $C$ for different $P_{GC}$ levels are
respectively shown in Figs.~\ref{fig:bubbleLengths_params}(a), \ref{fig:bubbleLengths_params}(b) and \ref{fig:bubbleLengths_params}(c).
Both the stretched exponent $\beta_l$ and the characteristic length $\kappa$ increase linearly with GC content
[Figs.~\ref{fig:bubbleLengths_params}(a) and \ref{fig:bubbleLengths_params}(b) respectively], with the PBD values being
always larger than the ones seen for the ePBD model. From Fig.~\ref{fig:bubbleLengths_params}(c) we see that the coefficient
$C$ exhibits for both the PBD (blue circles) and the ePBD (empty orange squares) systems an exponential decrease with
the GC content, capturing the overall decrease in the number of observed bubbles as $P_{GC}$ increases
(Fig.~\ref{fig:bubbleLengths}). The particular exponential fits of the preexponential factor are shown in Fig.~\ref{fig:bubbleLengths_params}(c) for the two models (blue solid curve for the PBD and the orange dashed curve for ePBD).
The difference between the $C$ values for the two models becomes larger for small $P_{GC}$ percentages,
with the ePBD values being always higher, in accordance to the larger $P_L(l)$ values observed for this model in Fig.~\ref{fig:bubbleLengths}.
Since for pure GC sequences ($P_{GC}=100\%$) both models exhibit similar $P_L(l)$ distributions in Figs.~\ref{fig:bubbleLengths}(a)
and \ref{fig:bubbleLengths}(b), the fitting parameters of Eq.~\eqref{eq:stretchFitLength} converge for
$P_{GC}=100\%$ in Fig.~\ref{fig:bubbleLengths_params}, as expected, while they are distinctly different in the other $P_{GC}$ cases.

\begin{figure}
    \centering
    \includegraphics[width=0.48\textwidth]{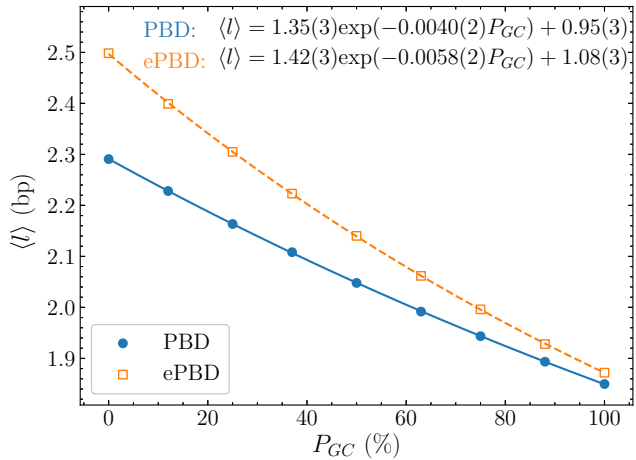}
    \caption{The average bubble length $\langle l \rangle$ as a function of the GC percentage $P_{GC}$ for the PBD
(blue circles) and ePBD (empty orange squares) models. Solid (dashed) curve shows fitting of the data with the exponential
function reported in the figure for the PBD (ePBD) model.}
    \label{fig:averageLength}
\end{figure}

The average bubble length $\langle l \rangle$ can  be computed as the number of base pairs in bubbles divided by the total number of bubbles \cite{Ares2007}:
\begin{equation}
    \label{eq:averageLength}
    \langle l \rangle = \frac{\sum_l l P_L(l)}{\sum_l P_L(l)}.
\end{equation}
Using the numerical results presented in Figs.~\ref{fig:bubbleLengths}(a) and \ref{fig:bubbleLengths}(b) and
Eq.~\eqref{eq:averageLength} we compute $\langle l \rangle$ for both the PBD and the ePBD models for various $P_{GC}$ percentages.
The obtained average bubble lengths are  shown in Fig.~\ref{fig:averageLength} by blue circles for the PBD model and by
empty orange squares for the ePBD system. These results indicate that the ePBD model exhibits generally longer average
bubble lengths than the PBD system for any GC percentage, once again in agreement with the findings of
Ref.~\cite{Alexandrov2009ePBD} that base pair openings tend to be larger in the ePBD model. The fine sequence dependence of the ePBD model through the stacking energy variation also demonstrates
greater sensitivity to the GC content of DNA, as its range of $\langle l \rangle$ values is wider, corresponding to
the longer tails seen in the bubble length distributions $P_L(l)$ in Fig.~\ref{fig:bubbleLengths}(b) for AT-rich sequences.
For both models we see an exponential decrease in $\langle l \rangle$ with increasing $P_{GC}$ values, which has been also
observed previously for the PBD model at physiological~\cite{Ares2007} and other temperatures~\cite{Kalosakas2009}.
Comparing our PBD results to the previous findings at the same temperature~\cite{Ares2007}, we see that while the average bubble length $\langle l \rangle$ for homogeneous AT sequences ($P_{GC}=0\%$) are the same in both investigations, in our study we find
longer average bubble lengths for GC-rich sequences. The former observation suggests that the larger threshold used
in Ref.~\cite{Ares2007} does not affect so much the average bubble lengths, but most likely the latter difference
is due to the base pair specific thresholds for openings used here as compared to a uniform threshold value.
%%%%%%%%%%%%%%%%%%%%%%
%%%%%%%%%%%%%%%%%%%%%%

%=========================
\section{Conclusions}
\label{sec:conclusions}

We have studied in detail the distributions of bubble lifetimes and bubble lengths in the PBD and ePBD models of
double stranded DNA, using base pair specific physical thresholds for determining base pairs to be open,
based on the consistency of the considered openings with the melting behavior of both systems.
In particular, the characteristic length scale of the Morse potential, Eq.~\eqref{eq:Morse}, for AT and GC base pairs
yields an effective threshold, as it is in agreement with the requirement that 50\% of the base pairs are open at the melting temperature.

Implementing these thresholds and performing extensive MD simulations we computed the bubble lifetime distributions
$P_l(t)$ of DNA molecules for different bubble lengths $l$, for sequences with a variable GC content
(Fig.~\ref{fig:lifetimesDist}). A two-peaked distribution was found for the case of single-site openings
[Fig.~\ref{fig:lifetimesDist}(a)], while for bubbles of length $l=2$ base pairs or greater, a stretched exponential, Eq.~\eqref{eq:stretchedExpLife}, with exponent $\beta \gtrsim 1$ fits the distribution quite accurately.
The ePBD model predicts bubbles to be generally longer-lived than the PBD model.

Bubble length distributions $P_L(l)$ were also produced from our simulations (Fig.~\ref{fig:bubbleLengths}). We found that
these distributions are described by usual stretched exponential functions, Eq.~\eqref{eq:stretchFitLength},
for both models. Our results show that longer bubbles are more likely to appear in the ePBD model, particularly when the
sequences have a larger proportion of AT base pairs. The observation of longer in size and also longer-lived
bubbles in the ePBD model is related to the lower average stacking energy and the larger base pair displacements occurring
in the ePBD model as compared to the original PBD model.

The distributions of bubble lifetimes $P_l(t)$, Eq.~\eqref{eq:stretchedExpLife}, and bubble length $P_L(l)$,
Eq.~\eqref{eq:stretchFitLength}, obtained in our work, in combination with the results of Figs.~\ref{fig:lifetimeParams}
and \ref{fig:bubbleLengths_params}, can be used to estimate the occurrence probability for any bubble of length $l$ and
lifetime $t$ in a sequence of specified GC content, i.e.~a fixed $P_{GC}$ percentage. Our results indicate that inherent
long-lived bubbles with lifetimes of the order of ps are infrequent, at least in the framework of the
algorithm considered here where fluctuations of the bubble size denote starting off a new bubble.
Larger bubbles exhibit exponentially decaying lifetimes.

Prospective future investigations include detailed studies of bubble lifetime and length distributions at functional sites
in DNA promoters, using the thresholds proposed in Sect.~\ref{sec:bubble_opening_thresholds}, or investigating
the effect of the opening amplitude on bubble lifetimes. Similar investigations can also be carried out using Langevin dynamics,
in order to consider the effects of a noisy environment on the obtained distributions, as well as exploring
the possibilities for a more flexible bubble-tracking algorithm.

%=========================
\section*{Acknowledgments} % (fold)

M.~H. and Ch.~S.~acknowledge support by the National Research Foundation (NRF) of South Africa. G.~K.~and Ch.~S.~were supported by the Erasmus+/International Credit Mobility KA107 program. We thank the High Performance Computing facility of the University of Cape Town and the Center for High Performance Computing of South Africa for providing computational resources for this project.

%=========================
%_______________________________________________

\end{document}